\documentclass[aps,twocolumn,superscriptaddress]{revtex4}

\usepackage{amsmath,amssymb,graphicx}
\usepackage{color}

\graphicspath{{fig-ps/}}

\definecolor{yblue}{rgb}{0.06, 0.3, 0.57}
\usepackage[pdftex]{hyperref}
\hypersetup{colorlinks=true,linkcolor=yblue,citecolor=yblue,urlcolor=yblue}

\begin{document}

\title{On Two-Component Dark-Bright
  Solitons in Three-dimensional Atomic Bose-Einstein Condensates}

\author{Wenlong Wang}
\email{wenlongcmp@gmail.com}
\affiliation{Department of Physics and Astronomy, Texas A$\&$M University,
College Station, Texas 77843-4242, USA}

\author{P.G. Kevrekidis}
\email{kevrekid@math.umass.edu}
\affiliation{Department of Mathematics and Statistics, University of Massachusetts,
Amherst, Massachusetts 01003-4515 USA}

\begin{abstract}
  In the present work, we revisit two-component Bose-Einstein condensates
  in their fully three-dimensional (3d) form. Motivated by earlier studies
  of dark-bright solitons in the 1d case, we explore the stability
  of these structures in their fully 3d form in two variants. In
  one the dark soliton is planar and trapping a planar bright (disk)
  soliton. In the other case, a dark spherical shell soliton creates
  an effective potential in which a bright spherical shell of atoms
  is trapped in the second component.
  We identify these solutions as numerically exact states (up to
  a prescribed accuracy) and perform a Bogolyubov-de Gennes
  linearization analysis that illustrates
  that both structures can be dynamically stable in suitable intervals
  of sufficiently low chemical potentials. We corroborate this finding
  theoretically by analyzing the stability via degenerate perturbation
  theory near the linear limit of the system.
  When the solitary waves are found to be unstable, we explore their dynamical
  evolution via direct numerical simulations which, in turn, reveal
  novel waveforms that are more robust. Finally, using the SO$(2)$
  symmetry of the model, we produce multi-dark-bright planar or shell
  solitons involved in pairwise oscillatory motion.
\end{abstract}

\maketitle

\section{Introduction}

Multi-component nonlinear waves have been explored in a variety
of physical contexts~\cite{kivshar,stringari,rip}. Some of most
prototypical ones among them, however, have been nonlinear
optics~\cite{kivshar}, as well as atomic Bose-Einstein
condensates (BECs)~\cite{stringari,rip}. A relevant model
in both cases is the one proposed by Manakov~\cite{manakov,intman1},
which has also been a theme of intense focus within mathematical
physics due to its integrability~\cite{ablowitz}. This has
generated the possibility of producing bright vector solitons
in the case of self-focusing nonlinearities (in optics; self-attractive
interactions in the atomic realm), and dark vector
solitons for self-defocusing ones (self-repulsive interactions
in the atomic case). In the latter case, also more exotic
states such as the so-called dark-bright solitary waves~\cite{siambook}
can be generated.
Higher-dimensional variants of the latter will be a focal point
of the present study.

Dark-bright solitary waves emerge in self-defocusing systems, although bright
solitons generally are not possible in them, for a simple reason: dark
solitons form a waveguiding potential that traps the light or matter
(in optics or BECs, respectively) in the second component~\cite{rip}. 
The robustness of the resulting excitations enabled the pioneering
observation of such states in photorefractive media in the works
of~\cite{seg1,seg2}. More recently, motivated by their theoretical
prediction in an atomic (trapped) setting by~\cite{buschanglin},
a substantial number of experimental explorations emerged
in the atomic realm~\cite{hamburg,pe1,pe2,pe3,pe4,pe5,azu}. These addressed
a diverse array of questions, including their ways of
generation, e.g., via phase imprinting and counterflow
among others~\cite{hamburg,pe1,pe4}, effects of dimensionality~\cite{pe2},
their bound-state pairs~\cite{pe3,lia}, or interactions with external
potentials~\cite{azu}. In parallel, to these, variants of the
states such as the so-called dark-dark solitons arising due to the
invariance of the Manakov model to SO$(2)$ rotation~\cite{pe4,pe5}
have also been extensively examined.

Most of the above mentioned studies on dark-bright solitary
waves have been experimentally performed in setups that were
higher than one-dimensional (1d) --although often the trapping was
such that the system could be deemed to be, to a reasonable
approximation, in a 1d regime. On the other hand, much of
the theoretical analysis and especially the analytically available
exact solutions~\cite{christo,vdbysk1,vddyuri,ralak,dbysk2,shepkiv,parkshin}
are found in 1d settings. This naturally prompts the question:
are there analogues of dark-bright solitons in genuinely higher
dimensional settings and, if so, can they be stable --possibly
in some appropriate parametric regimes? Our aim in the present
work is to indicate that the answer to both questions can
be positive, in fact, for multiple different installments
of dark-bright solitary waves.

In particular, we examine one variant of a dark-bright soliton
where the dark soliton is genuinely one-dimensional i.e., it
represents a hyperbolic tangent solution along one of the systems
axis, being homogeneous in the other two directions. This creates
a dark soliton ``plane''. The resulting planar effective potential, in the
spirit of the standard dark-bright soliton, can trap a bright
mass of light/atoms i.e., a planar bright soliton. We also explore
a more three-dimensional installment of the relevant waveform
motivated by the recent illustration of dark spherical shell
(DSS) solitons~\cite{dss}, in which a DSS traps through its density
dip a bright shell of atoms in the second component. In both cases,
we examine these states in the presence of parabolic trapping
as is customary in the realm of atomic BECs~\cite{stringari,rip,siambook}.
Remarkably, we find that both of these patterns, both the
planar dark bright (PDB) and the spherical dark bright (SDB)
can be dynamically stable for suitable choices of chemical potentials.
Nevertheless, as may be expected from their one-component analogues,
both states have destabilization mechanisms that arise via
a set of pitchfork (symmetry breaking) bifurcations that we
analyze. These bifurcations are found to be interesting in their
own right as they lead to the emergence of novel states (ones
with multiple vortex-line-bright or vortex-ring-bright
solitons~\cite{stathis}) that are worthwhile to
explore further. Hence, the usefulness of the present study
extends to the identification of the latter states as well.
Lastly, we utilize the SO$(2)$ symmetry of
the Manakov model to produce variants of the PDB and SDB
(in the same vein as the symmetry is used to produce dark-dark
solitons from dark-bright ones~\cite{pe4,pe5}).

Our presentation of the above features will be structured as
follows. In Sec.~\ref{setup} we present the theoretical and numerical
setup of the system and the associated methods of analysis.
In Sec.~\ref{results}, we present our systematic numerical results and
comparison with theoretical predictions. Finally, our conclusions and a
number of open problems for future consideration are given in
Sec.~\ref{conclusion}.





\section{Model and Computational Setup}
\label{setup}

\subsection{The Gross-Pitaevskii equation}

In the framework of mean-field theory, and for sufficiently
low-temperatures, the dynamics of a two-component 3d repulsive BECs, 
confined by a time-independent trap $V$, is described by the following
dimensionless Gross-Pitaevskii equation (GPE) [see~\cite{siambook,emergent}
  for relevant reductions to dimensionless units]

\begin{eqnarray}
i \frac{\partial \psi_1}{\partial t} &=& -\frac{1}{2} \nabla^2 \psi_1+V \psi_1 +(| \psi_1 |^2+| \psi_2 |^2) \psi_1  \nonumber \\
i \frac{\partial \psi_2}{\partial t} &=& -\frac{1}{2} \nabla^2 \psi_2+V \psi_2 +(| \psi_1 |^2+| \psi_2 |^2) \psi_2,
\end{eqnarray}
where $\psi_1(x,y,z,t)$ and $\psi_2(x,y,z,t)$ are the macroscopic wavefunctions.
We have in mind in the present setting a scenario of equal masses
associated with the hyperfine states of the same gas
(e.g.,~$^{87}$Rb)~\cite{rip}.
It is in that vein that we consider it to be a good approximation
to assume that the scattering length ratios (for intercomponent
and intracomponent interactions) are near unity~\cite{opanchuk}.
Furthermore, we examine the typical (experimentally) case
ofa harmonic trap of the form: 
\begin{equation}
V=\frac{1}{2} \omega_{\rho}^2 \rho^2+\frac{1}{2} \omega_z^2 z^2,
\label{potential}
\end{equation}
where $\rho=\sqrt{x^2+y^2}$, $\omega_{\rho}$ and $\omega_z$ are the
trapping frequencies along the $x\textrm{-}y$ plane and the vertical direction $z$, respectively.
Note that the potential has rotational symmetry with respect to the $z$-axis.
In our numerical simulations, we focus on the isotropic (spherical)
case with
$\omega=\omega_{\rho}=\omega_z=1$.

In this system, we seek stationary states of the form
\begin{eqnarray}
\psi_1(\vec{r},t) &=& \psi^0_1(\vec{r})e^{-i\mu_1t} \nonumber \\
\psi_2(\vec{r},t) &=& \psi^0_2(\vec{r})e^{-i\mu_2t}
\end{eqnarray}
which, in turn, lead to the stationary equations:
\begin{eqnarray}
\label{SS1}
-\frac{1}{2} \nabla^2 \psi^0_1+V \psi^0_1 +(| \psi^0_1 |^2+| \psi^0_2 |^2) \psi^0_1 &=& \mu_1 \psi^0_1 \nonumber \\
-\frac{1}{2} \nabla^2 \psi^0_2+V \psi^0_2 +(| \psi^0_1 |^2+| \psi^0_2 |^2) \psi^0_2 &=& \mu_2 \psi^0_2,
\end{eqnarray}

where $\mu_1$ and $\mu_2$ are the chemical potentials of the first and second components, respectively. A theoretical analysis of the existence
of solutions to Eqs.~(\ref{SS1}) can be presented in the vicinity
of the vanishing amplitude limit where the problem becomes
linear (and hence analytically tractable). We explore a perturbative
analysis near this limit next.

\subsection{The degenerate perturbation theory}
The eigenstates and eigenvalues of Eqs.~(\ref{SS1})
in the linear limit are the eigenstates of the
3d quantum harmonic oscillator. We now seek to identify how these eigenstates
and eigenvalues are modified by the presence of the nonlinearity
as we depart from this limit. 
Near the  limit, we apply the expansion for the
eigenstates and eigenvalues as follows:
\begin{eqnarray}
\psi_1^0 &=& \sqrt{\epsilon}\psi_{10}+\epsilon^{3/2} \psi_{11} + \dots, \nonumber \\
\psi_2^0 &=& (\sqrt{\epsilon \eta}\psi_{20}+(\epsilon \eta)^{3/2} \psi_{22})
 \dots,
\nonumber \\
\mu_1 &=& \mu_{10}+\epsilon \mu_{11} + \dots,
\nonumber \\
\mu_2 &=& \mu_{20}+\epsilon \eta \mu_{22} + \dots,
\label{expansion}
\end{eqnarray}
where $\psi_{10},\psi_{20},\mu_{10}$ and $\mu_{20}$ are normalized simple harmonic oscillator eigenstates and eigenenergies of the first and second components, respectively; $\mu_{11}$ and $\mu_{22}$ are two parameters that need to be determined self-consistently using Eqs.~(\ref{SS1}). The parameter
$\eta$ represents the ratio of the perturbations from the linear limit of the two components in the units of $\mu_{11}$ and $\mu_{22}$, respectively. $\eta$ depends on the selected path in the $(\mu_1,\mu_2)$ plane. We use a linear path and $\eta$ is $O(1)$ in our simulations.

To $O(\epsilon^{3/2})$, we get as a self-consistency condition
from Eqs.~\ref{SS1}:
\begin{eqnarray}
\mu_{11} &=& \int (|\psi_{10}|^2+\eta |\psi_{20}|^2) |\psi_{10}|^2 d^3x,
\nonumber \\
\mu_{22} &=& \frac{1}{\eta} \int (|\psi_{10}|^2+\eta |\psi_{20}|^2) |\psi_{20}|^2 d^3x.
\end{eqnarray}

We then define the slope of a linear path $s=\frac{\mu_1-\mu_{10}}{\mu_2-\mu_{20}}$ in the $(\mu_1,\mu_2)$ plane, and $I_{ij}=\int |\psi_{i0}|^2|\psi_{j0}|^2 d^3x$, and note that $I_{12}=I_{21}$, then

\begin{equation}
\eta = \frac{I_{11}-sI_{12}}{sI_{22}-I_{12}}.
\end{equation}

Next, we use this decomposition in order to identify the linear
stability eigenvalues of perturbations around the equilibria
of Eqs.~(\ref{SS1}). In this, so-called, Bogolyubov-de Gennes (BdG)
analysis, we utilize the perturbation ansatz
\begin{eqnarray}
  \psi_1 &=& e^{-i \mu_1 t} \left[ \psi_1^0 + \delta
  \left(a(\vec{r}) e^{\lambda t} + b^{\star}(\vec{r}) e^{\lambda^{\star} t}
  \right) \right]
  \label{pert1}
  \\
  \psi_2 &=& e^{-i \mu_2 t} \left[ \psi_2^0 + \delta
  \left(c(\vec{r}) e^{\lambda t} + d^{\star}(\vec{r}) e^{\lambda^{\star} t}
  \right) \right]
  \label{pert2}
\end{eqnarray}
and derive, to O$(\delta)$ the linearized equations
$\lambda (a,b,c,d)^T=M (a,b,c,d)^T$ where the eigenvalues
are denoted by $\lambda$, the eigenvectors by
$(a,b,c,d)^T$ ($^T$ stands for transpose) and the linearization
matrix $M$ is given in Appendix A.

In the matrix $M$,
if we separate the diagonal terms $-\frac{1}{2} \nabla^2 +V -\mu_1$ into $-\frac{1}{2} \nabla^2 +V -\mu_{10} - (\mu_1-\mu_{10})$ and $-\frac{1}{2} \nabla^2 +V -\mu_2$ into $-\frac{1}{2} \nabla^2 +V -\mu_{20} - (\mu_2-\mu_{20})$, the eigenvalues and eigenstates of the diagonal parts with $-\frac{1}{2} \nabla^2 +V -\mu_{10}$ and $-\frac{1}{2} \nabla^2 +V -\mu_{20}$ are known, and the rest of the terms are all $O(\epsilon)$ (upon the expansion of Eqs.~(\ref{expansion}))
hence they  can be treated as perturbations. Therefore, one can apply the degenerate perturbation analysis order-by-order so as to identify
the eigenvalues and eigenstates of the $M$ matrix.
The basis we use consists of the simple harmonic oscillator states in the first, second, third and fourth elements of the $4\times1$ vector space that we can call up-up states, up-down states, down-up states and down-down states. For example,
for a simple harmonic oscillator state $|mnp\rangle$ with eigenvalue $E_{mnp}$, the four states are as ($|mnp\rangle,0,0,0)^T$, (0,$|mnp\rangle,0,0)^T$, ($0,0,|mnp\rangle,0)^T$ and ($0,0,0,|mnp\rangle)^T$, with eigenvalues $E_{mnp}-\mu_{10}$, $\mu_{10}-E_{mnp}$, $E_{mnp}-\mu_{20}$, $\mu_{20}-E_{mnp}$, with a factor of $-i$.

The principal nonlinear states that we consider in what follows
are the PDB and SDB solitary waves. The former one (PDB) at the
linear limit corresponds to a $\psi_{10}=|001\rangle$ state in
the first component. This state,
as nonlinearity increases, has been shown to morph
into a planar dark soliton,
as was discussed in~\cite{russell}. This is coupled for the PDB
with the ground state $\psi_{20}= |000\rangle$ in the second component.
On the other hand, for the SDB, as discussed in~\cite{dss}, the
linear analogue of dark spherical shell is given by:
\begin{eqnarray}
|\psi_{\rm{DSS}}\rangle_{\rm{linear}}&=&
\frac{1}{\sqrt{3}}\left(|200\rangle+|020\rangle+|002\rangle \right)
\nonumber
\\
&\propto& \left(\omega r^2-\frac{3}{2}\right) e^{-\omega r^2/2}.
\end{eqnarray}
Hence, we pick for the first component $\psi_{10}=\psi_{\rm{DSS}}$,
while for the second component once again we select the ground
state $\psi_{20}= |000\rangle$.

Then, within the realm of the degenerate perturbation theory (DPT),
the relevant states for eigenvalues of the PDB and the SDB solitary
waves near the spectrum of Im($\lambda$)=0, 1 and 2 (in units of
the trapping frequency) are listed in Table~\ref{para}.
For the corresponding states, the O$(1)$ prediction of the DPT
is Im($\lambda$)=0, 1 and 2, respectively, while the O$(\epsilon)$
prediction yields how these modes depart from the linear limit
(in a way proportional to the perturbation size $\epsilon$).
This prediction will be compared with the full numerical results
from the linearization of the matrix $M$ in Sec.~\ref{results}.

\begin{table*}
\caption{
Relevant states for eigenvalues of the planar dark bright (PDB) and spherical dark bright (SDB) solitons near the spectrum of Im($\lambda$)=0, 1 and 2 for the degenerate perturbation theory. Here $|mnp \rangle$ stands for eigenstates with all possible distinct permutations with the quantum numbers $m,n$ and $p$.
\label{para}
}
\begin{tabular*}{\textwidth}{c @{\extracolsep{\fill}} cccccc}
\hline
\hline
Im($\lambda$) &States &``up-up'' states  &``up-down'' states &``down-up'' states  &``down-down'' states \\
\hline
$0$ &PDB &$|001 \rangle$ &$|001 \rangle$  &$|000 \rangle$ &$|000 \rangle$ \\
\hline
$1$ &PDB  &$|002 \rangle$ $|011 \rangle$ &$|000 \rangle$  &$|001 \rangle$ & \\
\hline
$2$ &PDB  &$|003 \rangle$ $|012 \rangle$ $|111 \rangle$ &   &$|002 \rangle$ $|011 \rangle$ & \\
\hline
$0$ &SDB &$|002 \rangle$ $|011 \rangle$ &$|002 \rangle$ $|011 \rangle$ &$|000 \rangle$ &$|000 \rangle$ \\
\hline
$1$ &SDB  &$|003 \rangle$ $|012 \rangle$ $|111 \rangle$ &$|001 \rangle$ &$|001 \rangle$ & \\
\hline
$2$ &SDB  &$|004 \rangle$ $|013 \rangle$ $|022 \rangle$ $|112 \rangle$ &$|000 \rangle$ &$|002 \rangle$ $|011 \rangle$ & \\
\hline
\hline
\end{tabular*}
\end{table*}

A detailed analysis of the numerical method (and its use of the
symmetry of the solutions) is explained in Appendix B. For our
current purposes, suffice it to say that we obtain the full solution
of Eqs.~(\ref{SS1}) up to a prescribed tolerance via a 2d Newton
iteration (our solutions will not have a topological charge --and hence
a dependence on the azimuthal variables-- in the cylindrical coordinates used for the PDB and SDB).
Subsequently the BdG analysis is used, once again taking advantage
of the symmetry of the solution as is detailed in Appendix B, and the
numerical linearization spectrum of eigenvalues $\lambda$ is identified.
The process is repeated for our prescribed parametric variation over
$\mu$ and the resulting bifurcation as well as stability diagram
will be presented in comparison with the corresponding near-linear
theoretical prediction of the DPT.

\section{Numerical and analytical results}
\label{results}

We now turn to the numerical investigation of our higher-dimensional
dark-bright soliton states of interest.

\subsection{The PDB and SDB spectrum}

\begin{figure}[tb]
\begin{center}
  \includegraphics[width=7.5cm]{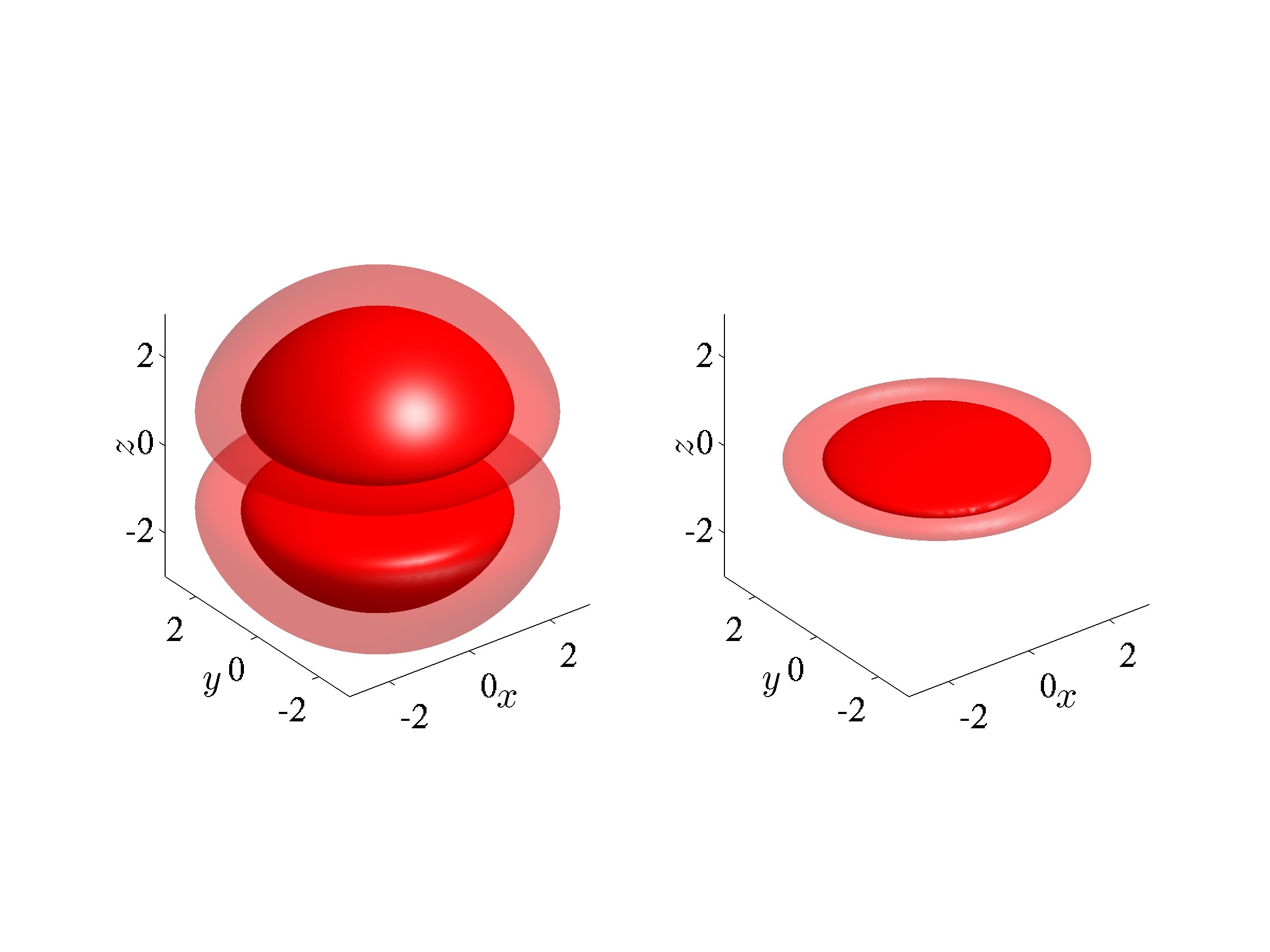}
  \includegraphics[width=7.5cm]{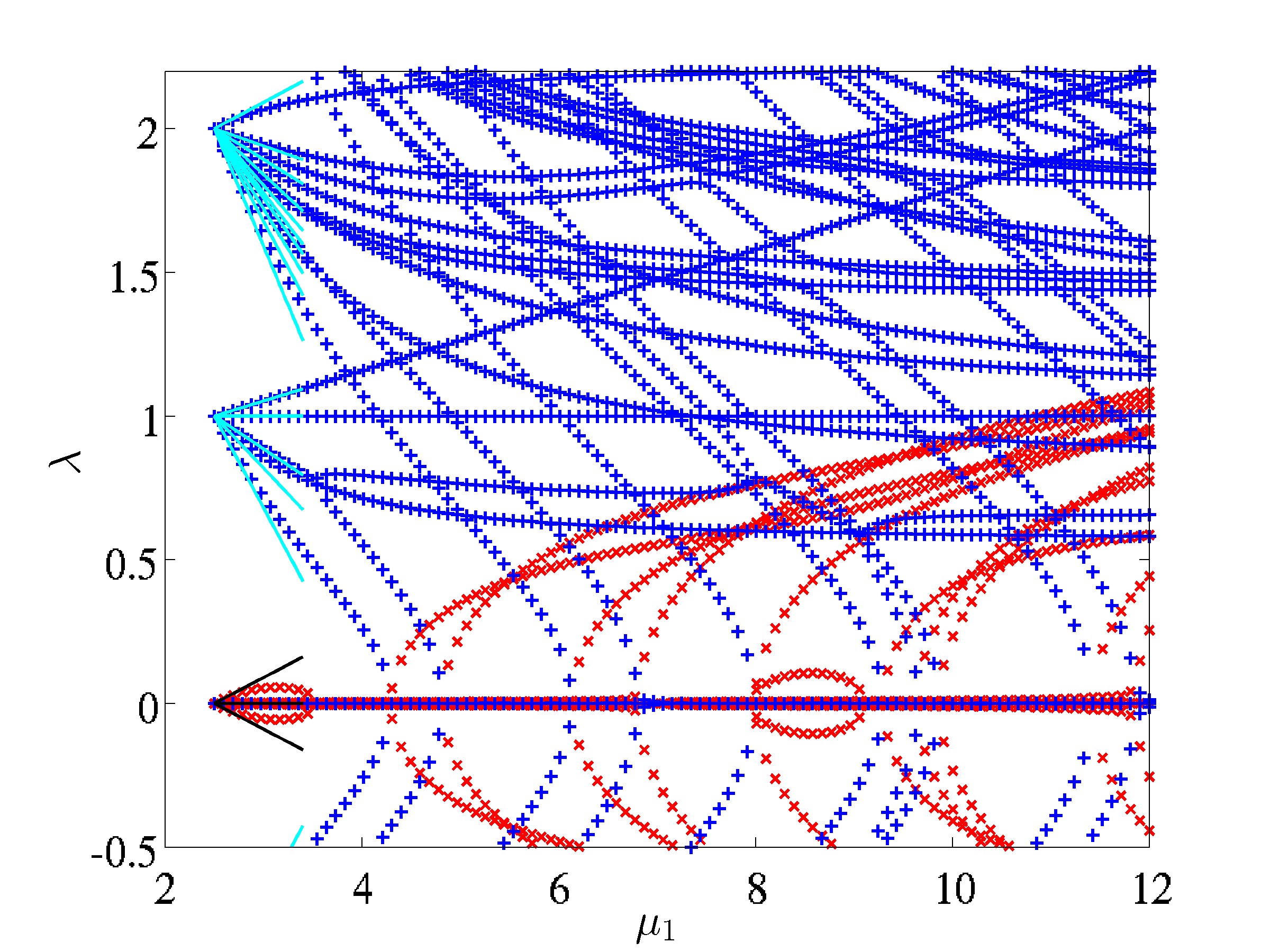}
\caption{
(Color online)
  Top panels: Density profile of a representative three-dimensional dark-bright soliton at $\mu_1=12$ and $\mu_2=10$.
  Bottom panel: Spectrum of the dark-bright soliton along the line from the linear limit to $\mu_1=12$ and $\mu_2=10$. The red crosses and blue pluses are unstable and stable eigenvalues, respectively. The black and cyan lines are unstable and stable eigenvalues calculated using the degenerate perturbation theory. Note that they agree well near the linear limit, and the dark-bright soliton has a small interval where it is completely stable near but not including the linear limit.
}
\label{DBS}
\end{center}
\end{figure}

\begin{figure}[tb]
\begin{center}
  \includegraphics[width=7.5cm]{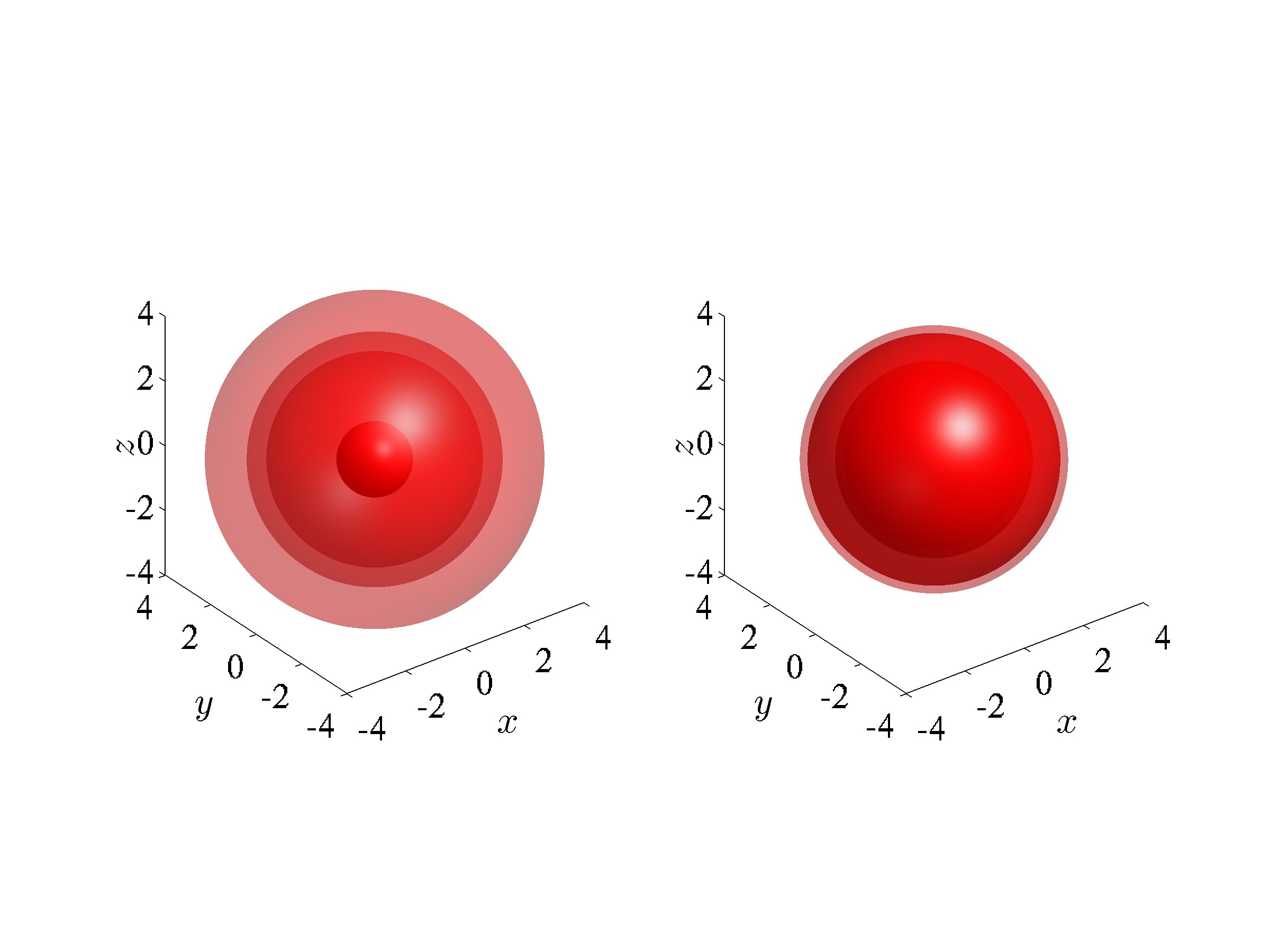}
  \includegraphics[width=7.5cm]{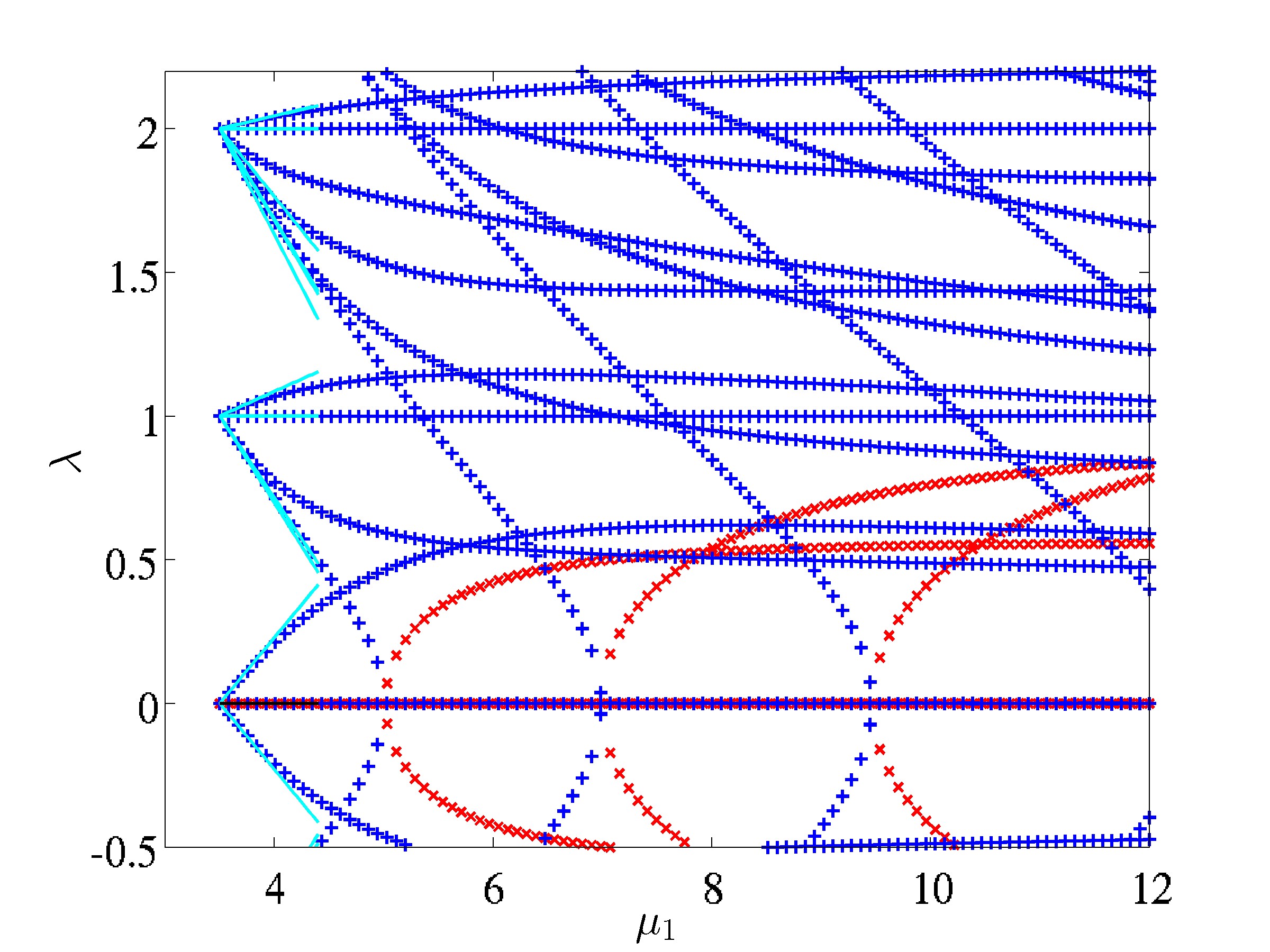}
\caption{
(Color online)
  Top panels: Density profile of a representative three-dimensional dark-shell-bright soliton at $\mu_1=12$ and $\mu_2=10$.
  Bottom panel: Spectrum of the dark-shell-bright soliton along the line from the linear limit to $\mu_1=12$ and $\mu_2=10$. The red crosses and blue pluses are unstable and stable eigenvalues, respectively. The black and cyan lines are unstable and stable eigenvalues calculated using the degenerate perturbation theory. Note that they agree well near the linear limit, and the dark-shell-bright soliton is remarkably stable near the linear limit.
}
\label{DSB}
\end{center}
\end{figure}

A typical stationary state and the spectrum of the PDB are shown in Fig.~\ref{DBS}, and similarly for the SDB in Fig.~\ref{DSB}.
The figures also illustrate a continuation over a
parametric line in the $(\mu_1,\mu_2)$ plane, in terms of the stability eigenvalues $\lambda$.
It is interesting that in the case $\mu_1 \approx \mu_2$, the qualitative
features of the spectrum are similar to the dark soliton and the dark soliton shell in one component BECs, respectively.
In particular, the planar DB similarly to the planar dark soliton is unstable
very near the linear limit (an instability captured quite accurately --as
are all eigenmodes near that limit-- by the degenerate perturbation theory).
On the other hand, the spherical DB similarly to the dark soliton shell
state of~\cite{dss} is stable in the immediate vicinity of the linear
limit, where once again the DPT fares very well in capturing the leading
order behavior of the associated eigenmodes. 
Nevertheless, as mentioned earlier, both states present chemical potential
intervals where the solitons are completely stable. In our simulations, we find they are stable in the intervals of $3.6<\mu_1<4.2$ and $3.5<\mu_1<4.9$ for PDB and SDB, respectively.
These spectral stability conclusions of the BdG analysis have been
confirmed through direct numerical computations using a 4th order
Runge-Kutta scheme.

\begin{figure}[tb]
\begin{center}
\includegraphics[width=7.5cm]{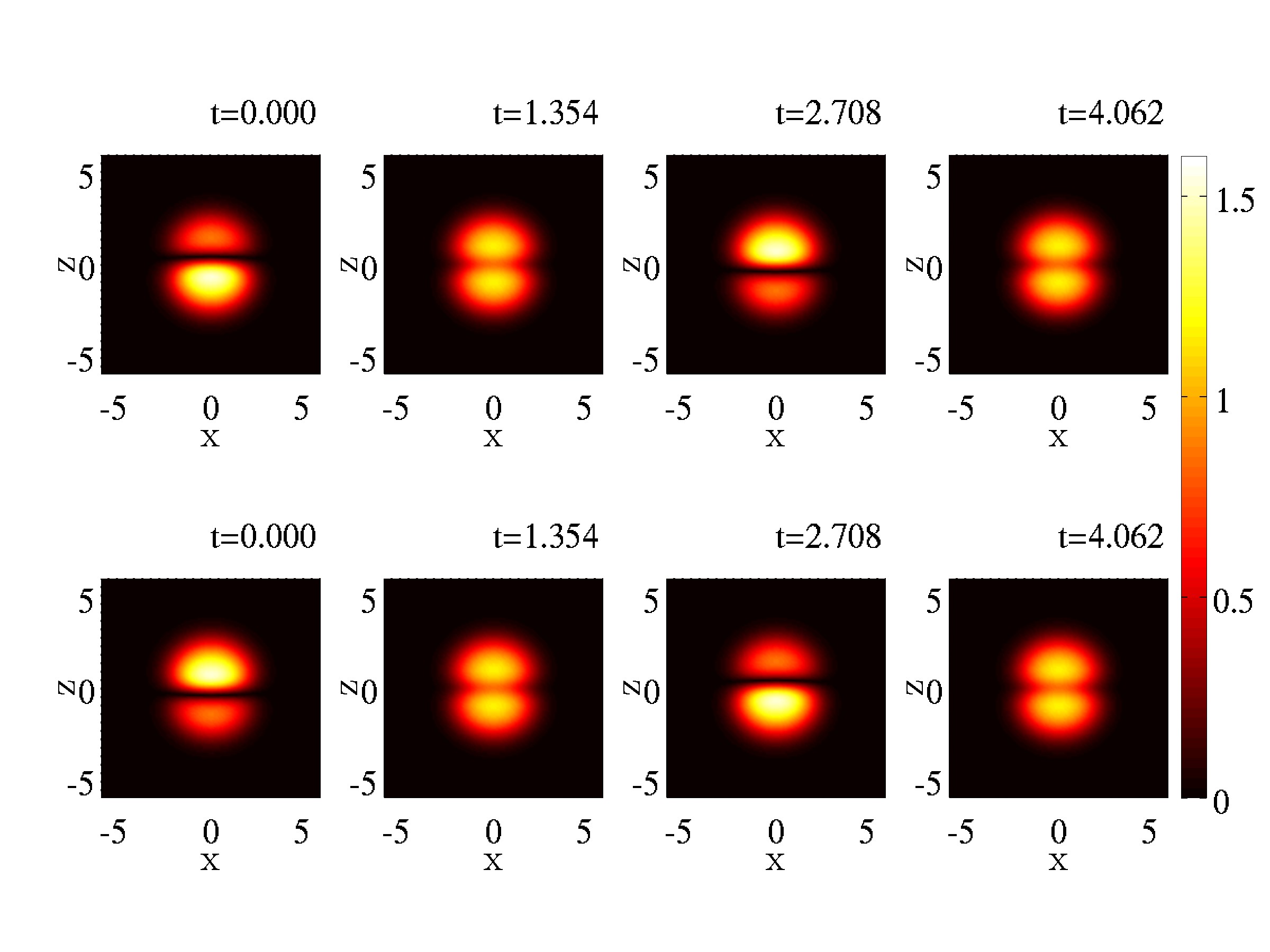}
\includegraphics[width=7.5cm]{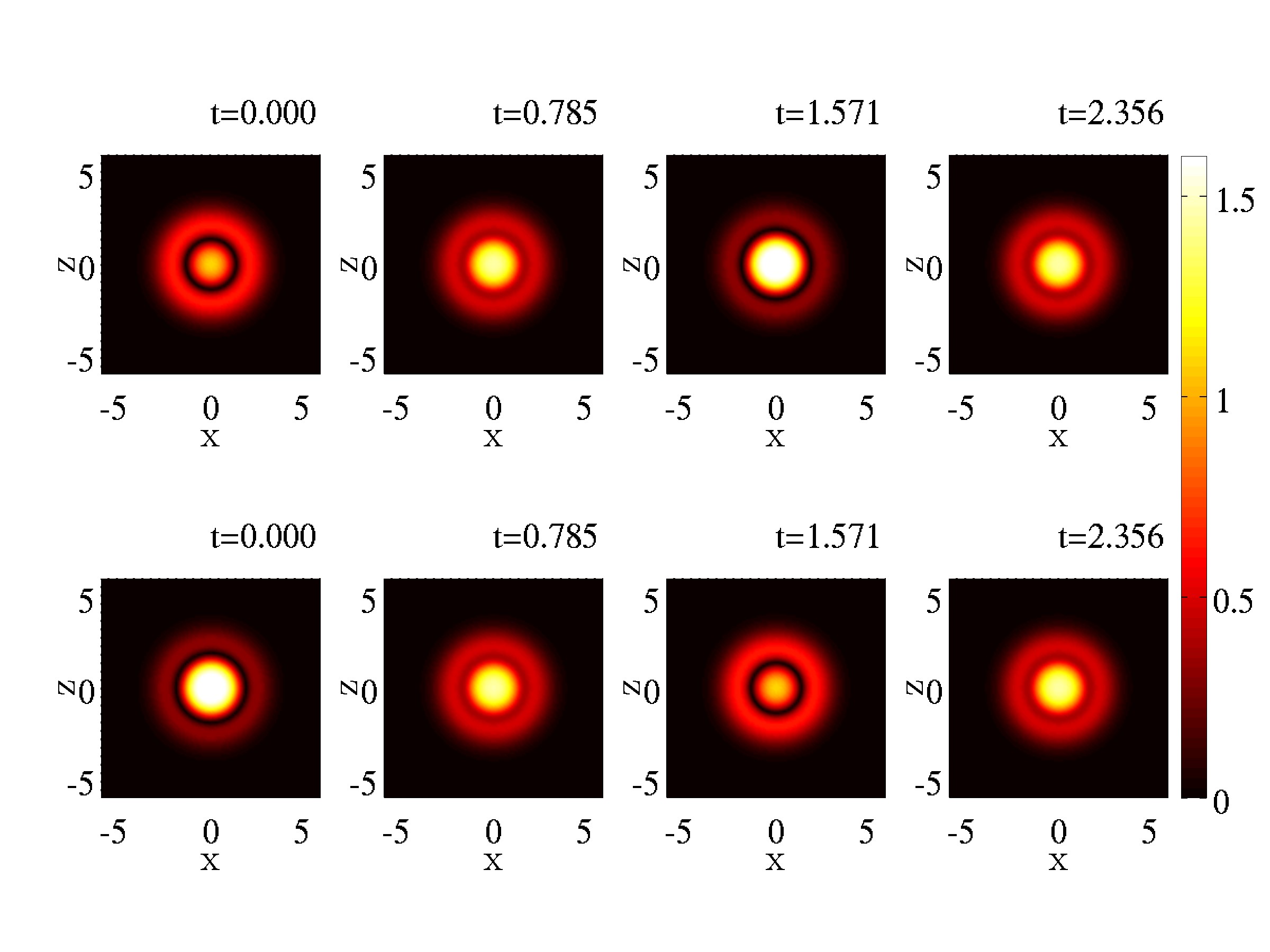}
\caption{
(Color online)
  Top two panels: Representative states of SO$(2)$ induced oscillation patterns at $t=0,\ T/4,\ T/2,\ 3T/4$ for $\mu_1=4.02$ and $\mu_2=2.86$ emerging
  from the
  rotation of the PDB. An oscillating pair of PDBs results.
  The oscillation period is $T=2\pi/(\mu_1-\mu_2) \approx 5.4165$. The dynamics is in good agreement with the theoretical prediction, and is checked to be robust up to $t=100$, showing indirectly the stability of the dark-bright soliton at the chemical potentials studied. Bottom two panels:
  Same thing for the SO$(2)$ rotation of the SDB, resulting in a pair
  of SDBs oscillating one around the other. 
  Here $\mu_1=4.86,\ \mu_2=2.86$, hence $T=\pi$.
}
\label{DBST}
\end{center}
\end{figure}

A complementary way to test this stability, but which also illustrates
a different type of pattern that can emerge in two-component atomic
BECs stems from the SO$(2)$ rotation-invariance of the Manakov model.
In particular any pseudo-spinor state $(\psi_1,\psi_2)$ can be
rotated by a unitary $2 \times 2$ matrix (here we use simply an orthogonal
rotation with an angle $\theta$),
resulting still in an exact solution of the model.
It is in this way that dark-dark solitons were obtained
from dark-bright ones in 1d and this justified their relevance
and spontaneous emergence in experiments~\cite{pe4,pe5}.
In the case of Fig.~\ref{DBST}, we observe the stable oscillation patterns
up to $t=100$ that result from the PDB (top) and SDB (bottom)
cases. We have studied both symmetric and asymmetric mixing of the two
components, and both are robust. Here we only show results of the symmetric
case, i.e., $\theta=\pi/4$. Representative states at $t=0,\ T/4,\ T/2,\ 3T/4$,
are shown where $T=2\pi/|\mu_1-\mu_2|$ is the oscillation period.
It can be seen that the SO$(2)$ rotation of the PDB yields a novel
state where there is a planar dark soliton in each component (one
of which is at $z>0$ and one at $z<0$, with a complementary bright
atomic mass in the other component). This can be dubbed
a PDB-PDB state. Similarly the SDB-SDB state of the bottom two panels,
resulting from the SO$(2)$ rotation of the SDB features an inner
and an outer SDB state with the inner and outer dark shells residing
in different components.
See Ref.~\cite{youtube} for dynamical movies illustrating the
details of the resulting oscillations.

\subsection{Instability Dynamics for the PDB}
We now turn to an examination of the unstable dynamics of
the two states of interest in the regions of chemical potential
where they become unstable. We can see that similarly to the
one-component structures
of~\cite{russell} (and~\cite{dss} for the SDBs), these states suffer
a cascade of pitchfork bifurcations leading to the emergence of a
sequence of unstable modes.

\begin{figure}[tb]
\begin{center}
\includegraphics[width=7.5cm]{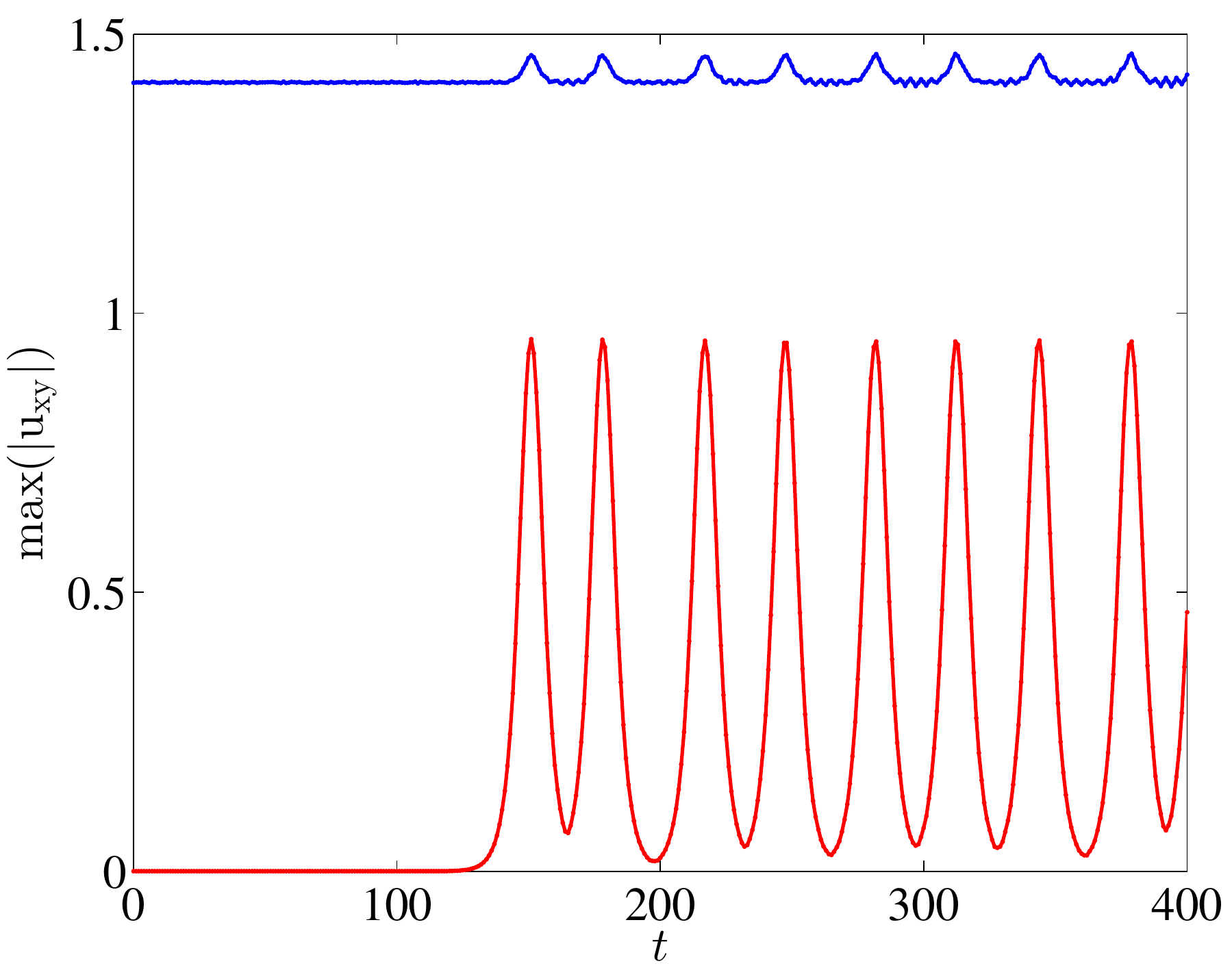}
\includegraphics[width=7.5cm]{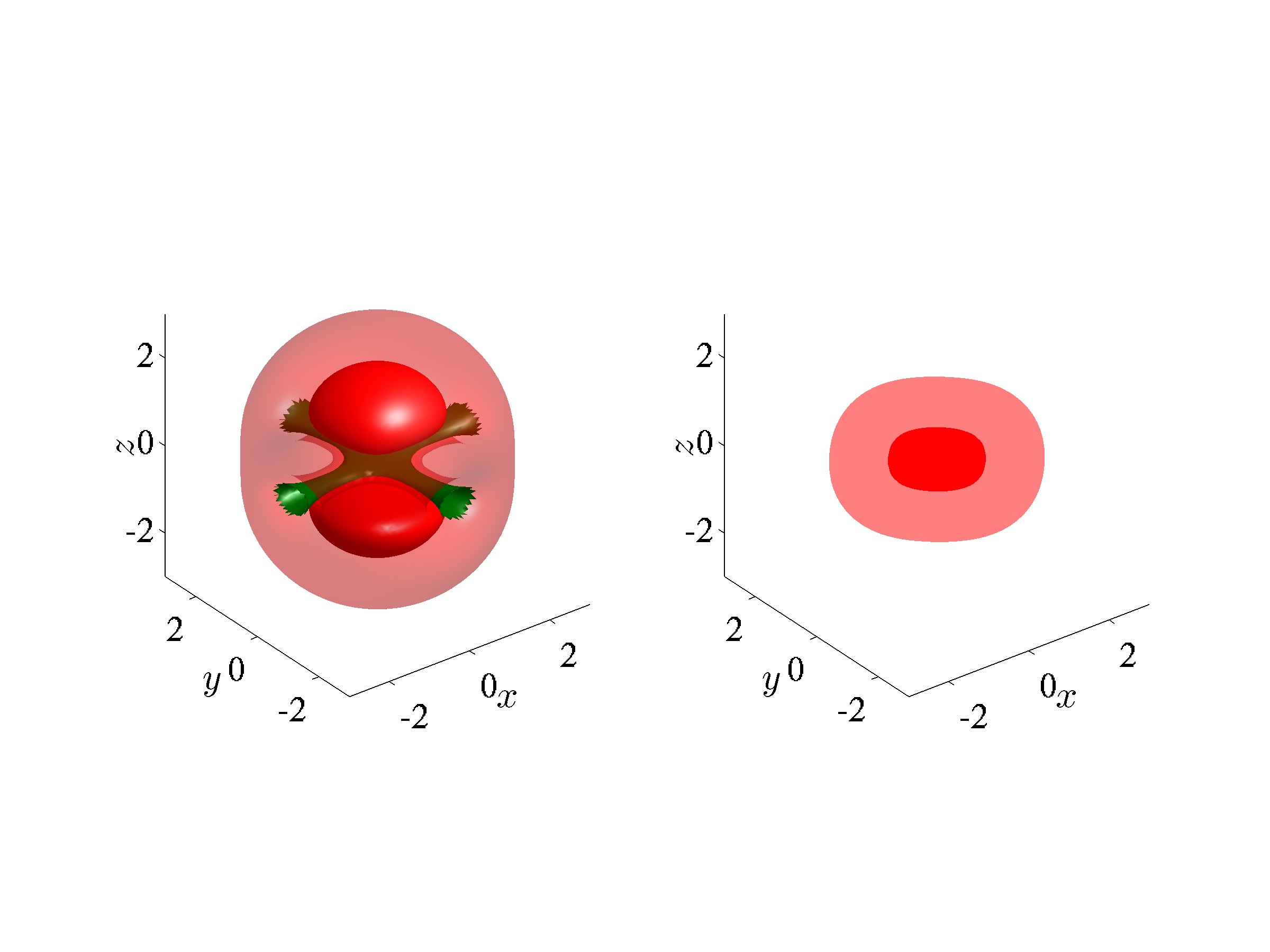}
\includegraphics[width=7.5cm]{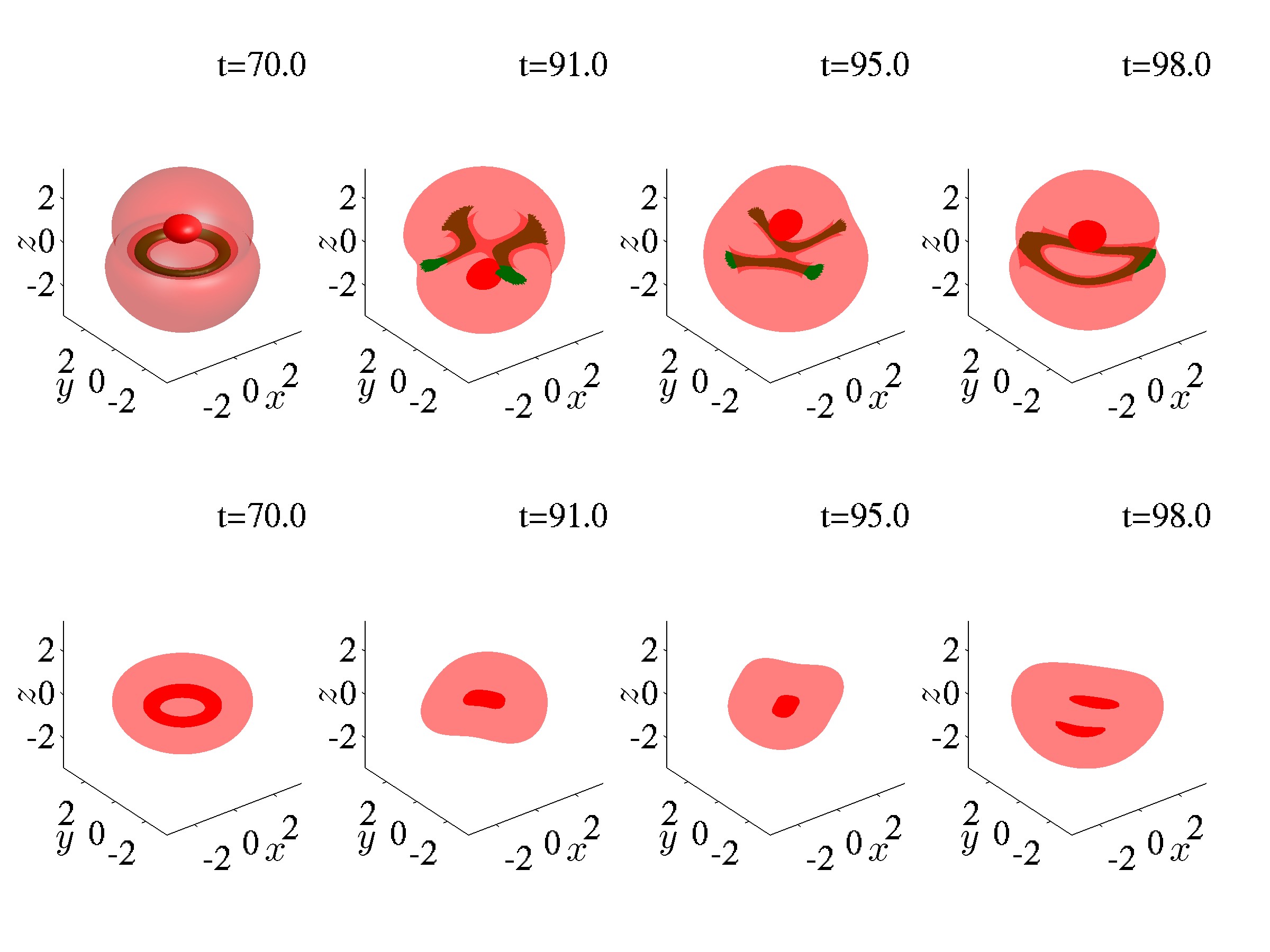}
\caption{
(Color online)
  Top panel: The maximum densities of the two components in the $x\textrm{-}y$ plane in the dynamics of the PDB at $\mu_1=4.685,\ \mu_2=3.455$ up to $t=400$. The condensates oscillate between the PDB and the 2VLB soliton.
  Second row: Density profile of a 2VLB state at $t=217$ in the above PDB dynamics.
  Bottom two rows: Dynamics of PDB with the instability mode
  leading to a VRB at $\mu_1=6.015$, $\mu_2=4.645$.}
\label{uxy3}
\end{center}
\end{figure}

The first unstable mode of a planar dark soliton has been shown to lead
to a 2 vortex line state; similarly, the unstable mode of a PDB in
a two-component setting yields a 2 vortex line state trapping respective
bright solitons i.e., a 2 vortex-line-bright (VLB) state.
We have confirmed this by exploring the dynamics at $\mu_1=4.685,\ \mu_2=3.455$ up to $t=400$. The top panel of Fig.~\ref{uxy3} shows the evolution of
this instability and how the waveform departs from the PDB leading
(upon its maximal deviation from it) to a 2 VLB state shown, a typical
example of which is shown in the middle panel of the figure for $t=217$.
However, then due to the recurrent dynamics of the Hamiltonian system,
it returns to the PDB, oscillating back and forth between these two
states over the time scale monitored.
Nevertheless, it is interesting that each time the 2VLB soliton forms,
it appears to forget its history and it possesses a different
orientation in the $x\textrm{-}y$ plane.

In the case of a planar dark soliton, a
second mode of instability (occurring very close to the emergence of
the 2 VL state) leads to the bifurcation of a vortex ring (VR).
Thus, we similarly expect that the second instability (after
then one near $\mu_1 \approx 4.3$ leading to 2 VLB) arising
near $\mu_1 \approx 4.8$
in Fig.~\ref{DBS} will lead to a vortex-ring-bright (VRB) solitonic state.
To examine the outcome of this instability, we have
monitored the dynamics at $\mu_1=6.015,\ \mu_2=4.645$ upto $t=200$. The VRB soliton was observed as shown in Fig.~\ref{uxy3}. We see that a VR is first formed, but the VR is unstable. It breaks through a quadrupolar ($n=2$)
Kelvin mode to a 2VL. Subsequently, the 2VL come closer, exchange their segments and hence change to a new orientation. The 2VL then reconnects to form a highly excited VR. The VR is seen to break anew to 2VL, reconnect and break again. The subsequent dynamics of the 2VL are quite complex, and can oscillate
within the trap.
Note that the bright component changes its shape in tune  with
the vortex lines and rings of the first component.


\subsection{Instability Dynamics for the SDB}
\begin{figure}[tb]
\begin{center}
\includegraphics[width=7.5cm]{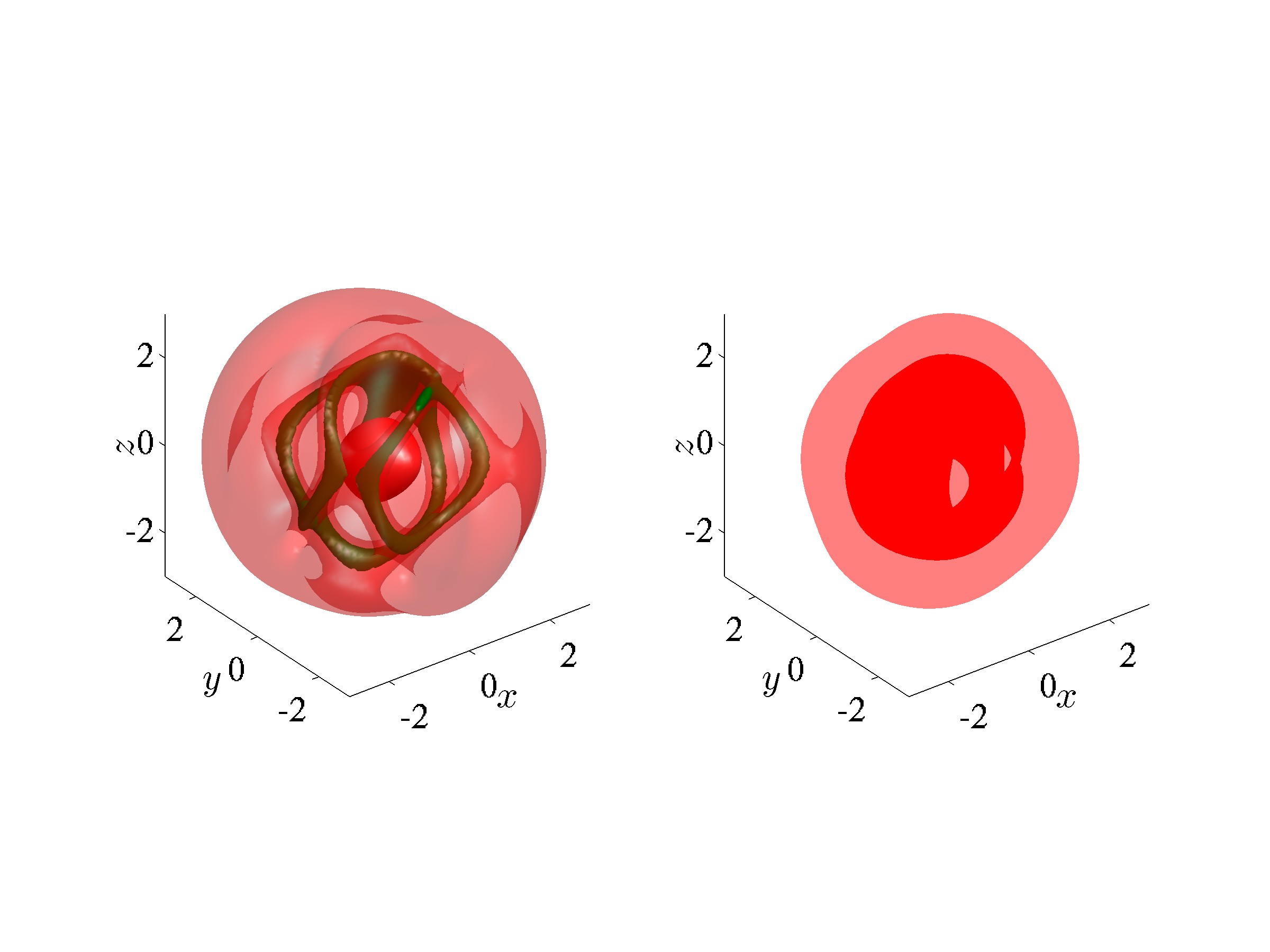}
\includegraphics[width=7.5cm]{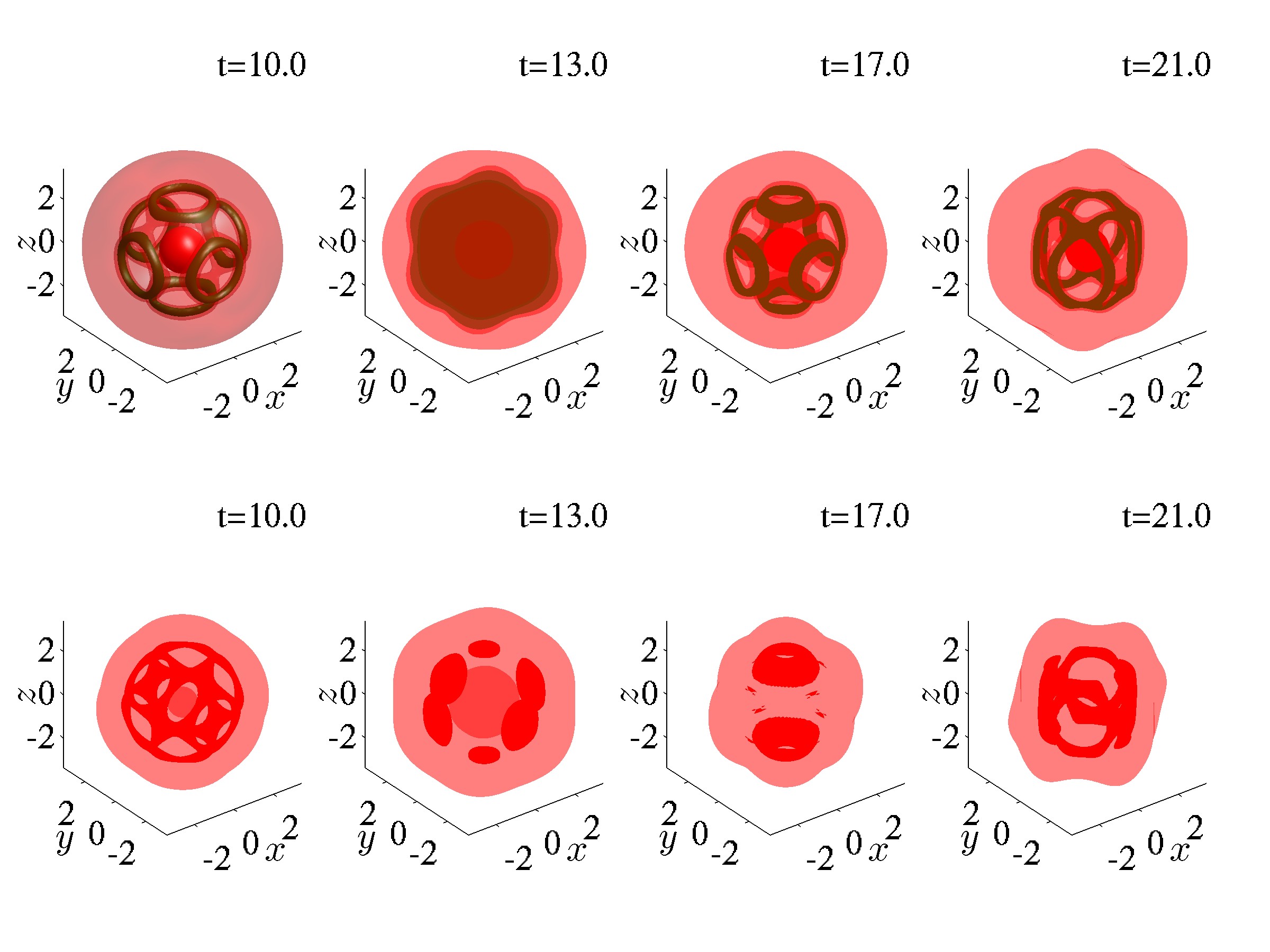}
\caption{
(Color online)
  Top panel: Density profile of a 6VLB state at $t=79$,
  resulting from the unstable dynamics of a SDB
  soliton at $\mu_1=6.9,\ \mu_2=4.9$.
  Bottom two rows: Dynamics of SDB at $\mu_1=9.28,\ \mu_2=7.28$
  leading through its instability mode to a 6VRB (the evolution
is shown at different times).}
\label{6VLB}
\end{center}
\end{figure}

Finally, we study the instability modes of the spherical dark
bright solitary wave. Similarly to the case of the PDB, we turn
to the analysis of the one-component state, in this case
the dark soliton shell of~\cite{dss} for insights on the dynamical
consequences of the modes of instability. We find there that
the first mode of instability of the one-component state leads
to a 6 vortex line ``cage'', while the second one to a  6 vortex ring cage.
We have thus followed the dynamics of our two-component SDB
state at $\mu_1=6.9,\ \mu_2=4.9$ as shown in
the top panel of Fig.~\ref{6VLB} (for $t=79$).
There, we observe indeed that the two-component analog of the 6 VL
state, the 6 VLB is emerging through the unstable dynamics of
the SDB. However, in turn,
this state  undergoes vortex line reconnections. For example, a pair of vertical vortex lines can exchange line segments and become horizontal.
The 6VL cage is also observed to reappear, yet its dynamics
leads to complex vortex line/ring dynamics and  reconnections which
we do not pursue further here.

In line with the above discussion, the second unstable
mode is expected to lead to a cage of vortex-ring-bright solitary
waves (a 6 VRB state). We have performed dynamical simulations
at $\mu_1=9.28,\ \mu_2=7.28$, and the 6 VRB cage is indeed
observed around $t=10$ as shown in the bottom panels of
Fig.~\ref{6VLB}. This state is more robust than the 6VLB cage,
yet it is subsequently deformed
for the chemical potentials studied.
Nevertheless, we observe its recurrence, as well as the formation
of more complex states involving VRBs in the relevant panels.


\section{Conclusions and Future Challenges}
\label{conclusion}

In this work, we considered generalizations of the notion of the
dark bright soliton in three-dimensional settings, including the planar
dark bright solitary wave and the (spherical) shell dark-bright
solitary wave. We constructed these structures systematically
from the low density, linear limit of the mean-field model and
continued them from there towards the high density, highly nonlinear
limit of the system. Although we are well aware of
the fact that for low densities, features such
as quantum fluctuations~\cite{carr} will be important, nevertheless
this limit is useful towards understanding how to systematically
construct such states and how to assess their stability characteristics.
In fact, in the vicinity of the linear limit, 
the developed (for the multi-component case) degenerate perturbation theory
enables a systematic characterization of the different eigenmodes
of the system.
On the numerical side, we utilized the symmetry of the solutions
to simplify the computation of the relevant eigenmodes which were
found to be in good agreement with the theory.
In the case of both the principal families considered,
intervals of stability of the solutions were identified, although
the states were found to be subject to symmetry breaking pitchfork
bifurcations. The latter were numerically identified
to lead to the emergence of vortex-line-bright and
vortex-ring-bright solitary waves (and also multi-structure
generalizations thereof). As an aside, we also explored
SO$(2)$ rotations of the families considered, leading to
pairs of planar and spherical dark bright states.


Our results lead to a number of interesting questions for future study.
We can see from this work that
vortex-line-bright solitons and vortex-ring-bright solitons
and multi-solitons may spontaneously
arise as a result of dynamical instabilities of other states. In that
light, these entities merit independent study in their own
right. The same is true for skyrmion states (consisting of a vortex
ring coupled to a vortex line). The latter have been studied dynamically
in earlier dynamical studies~\cite{ruosteko}, yet we propose a systematic
analysis of their spectrum. It would be especially interesting to
explore both for states such as the ones herein and for some of the
above proposed extensions whether in the highly nonlinear limit
a particle picture can be formulated describing the in-trap dynamics
and multi-state interactions, as has been done earlier for vortices
or vortex rings~\cite{siambook}. Such studies are presently in
progress and will be reported in future publications.

\begin{acknowledgments}

W.W.~acknowledges support from NSF-DMR-1151387.
P.G.K.~gratefully acknowledges the support of
NSF-DMS-1312856, NSF-PHY-1602994, as well as from
the ERC under FP7, Marie
Curie Actions, People, International Research Staff
Exchange Scheme (IRSES-605096).
The work of W.W. is supported in part by the Office of the Director of National Intelligence (ODNI), Intelligence Advanced Research Projects Activity (IARPA), via MIT Lincoln Laboratory Air Force Contract No.~FA8721-05-C-0002. The views and conclusions contained herein are those of the authors and should not be interpreted as necessarily representing the official policies or endorsements, either expressed or implied, of ODNI, IARPA, or the U.S.~Government. The U.S.~Government is authorized to reproduce and distribute reprints for Governmental purpose notwithstanding any copyright annotation thereon. We thank Texas A\&M University for access to their Ada and Curie clusters.

\end{acknowledgments}

\section*{APPENDIX A: LINEAR STABILITY CONSIDERATIONS}

Linear stability analysis (Bogolyubov-de Gennes
  Analysis) shows that the stability is governed by the eigenvalues $\lambda$ of the following 4$\times$4 matrix $M$:
\begin{eqnarray}
M_{11}&=&-i\left( -\frac{1}{2} \nabla^2  +V  -\mu_1+2|\psi^0_1|^2+|\psi^0_2|^2 \right), \nonumber \\
M_{12}&=&-i (\psi_1^{0})^2, \nonumber \\
M_{13}&=&-i\psi^0_1\psi_2^{0*}, \nonumber \\
M_{14}&=&-i\psi^0_1\psi^0_2, \nonumber \\
M_{21}&=&i(\psi_1^{0*})^2, \nonumber \\
M_{22}&=&i\left( -\frac{1}{2} \nabla^2 +V  -\mu_1+2|\psi^0_1|^2+|\psi^0_2|^2 \right), \nonumber \\
M_{23}&=&i\psi_1^{0*}\psi_2^{0*}, \nonumber \\
M_{24}&=&i\psi_1^{0*}\psi^0_2, \nonumber \\
M_{31}&=&-i\psi_1^{0*}\psi^0_2, \nonumber \\
M_{32}&=&-i\psi^0_1\psi^0_2, \nonumber \\
M_{33}&=&-i\left( -\frac{1}{2} \nabla^2 +V  -\mu_2+|\psi^0_1|^2+2|\psi^0_2|^2 \right), \nonumber \\
M_{34}&=&-i(\psi_2^{0})^2, \nonumber \\
M_{41}&=&i\psi_1^{0*}\psi_2^{0*}, \nonumber \\
M_{42}&=&i\psi^0_1\psi_2^{0*}, \nonumber \\
M_{43}&=&i(\psi_2^{0*})^2, \nonumber \\
M_{44}&=&i\left( -\frac{1}{2} \nabla^2 +V -\mu_2+|\psi^0_1|^2+2|\psi^0_2|^2 \right).
\end{eqnarray}
All the elements of the matrix aside from the diagonal ones are
multiplicative, while $\psi_{1,2}^{0}$ denotes the (two components
of the) stationary state
around which the linearization is performed.

\section*{APPENDIX B: THE PARTIAL-WAVE METHOD}

To compute the stability of stationary states away from the linear limit,
we employ numerical computations. This is however very demanding in the full 3d setting. For stationary states with rotational symmetry up to a topological charge, the work can be significantly reduced by working in the $\rho \textrm{-} z$ plane. It is straightforward to compute the stationary states,
taking advantage of the relevant symmetry. Here we present how to compute the stability for two-component solitons, generalizing the relevant discussion of Ref.~\cite{dss} for one-component solitons. Although we will utilize the method
for the real PDB and SDB profiles in $\psi_1^0$ and $\psi_2^0$, we present
the (partial-wave) method for the stability in a more general form.

The Laplacian in the cylindrical coordinates $(\rho,\phi,z)$ is decomposed as follows:
\begin{equation}
\nabla^2 f =\Delta_H f+\frac{\Delta_G f}{\rho^2},
\end{equation}
where its components read:
\begin{eqnarray}
\Delta_H f &=&\frac{1}{\rho} \frac{\partial}{\partial \rho} \left(\rho \frac{\partial f}{\partial \rho}\right)+\frac{\partial^2 f}{\partial z^2},\\
\Delta_G f &=&\frac{\partial^2 f}{\partial \phi^2}.
\end{eqnarray}

For states with rotational symmetry with topological charges $S_1$ and $S_2$, we have

\begin{eqnarray}
\psi^0_1(\vec{r}) &=& \psi^0_1(\rho,z)e^{iS_1\phi} \nonumber \\
\psi^0_2(\vec{r}) &=& \psi^0_2(\rho,z)e^{iS_2\phi},
\end{eqnarray}

and Eqs.~(\ref{SS1}) become:

\begin{eqnarray}
\label{SSP1}
-\frac{1}{2} \Delta_H \psi^0_1 + \frac{S_1^2}{2\rho^2}\psi^0_1 + V \psi^0_1 +(| \psi^0_1 |^2+| \psi^0_2 |^2) \psi^0_1 &=& \mu_1 \psi^0_1 \nonumber \\
-\frac{1}{2} \Delta_H \psi^0_2 + \frac{S_2^2}{2\rho^2}\psi^0_2 + V \psi^0_2 +(| \psi^0_1 |^2+| \psi^0_2 |^2) \psi^0_2 &=& \mu_2 \psi^0_2.
\nonumber \\
\end{eqnarray}

For a generic perturbation, using Fourier analysis,

\begin{eqnarray}
\label{Perturbation}
\psi_1 &=& e^{iS_1\phi} \left[ \psi^0_1+\delta \sum\limits_{m} \left[ a_{m}(\rho,z,t)e^{im\phi}+c_{m}^*(\rho,z,t)e^{-im\phi} \right] \right] \nonumber \\
\psi_2 &=& e^{iS_2\phi} \left[ \psi^0_2+\delta \sum\limits_{m} \left[ b_{m}(\rho,z,t)e^{im\phi}+d_{m}^*(\rho,z,t)e^{-im\phi} \right] \right],
\nonumber \\
\end{eqnarray}
Substituting this expansion into the GPE, keeping the terms to
O$(\delta)$ of the perturbations and using the uniqueness of the Fourier series of both sides, one can get for each mode $m$:

\begin{eqnarray}
i\dot{a}_m &=& -\frac{1}{2} \Delta_H a_m + \frac{(m+S_1)^2}{2\rho^2} a_m  +V a_m -\mu_1 a_m \nonumber \\
 &+& (2|\psi^0_1|^2+|\psi^0_2|^2) a_m +\psi^0_1 \psi_2^{0*} b_m + \psi_1^{02} c_m + \psi^0_1\psi^0_2 d_m \nonumber \\
i\dot{c}_m^* &=& -\frac{1}{2} \Delta_H c_m^* + \frac{(m-S_1)^2}{2\rho^2} c_m^*  +V c_m^* -\mu_1 c_m^* \nonumber \\
&+& (2|\psi^0_1|^2+|\psi^0_2|^2) c_m^*+ \psi_1^{02} a_m^* +\psi^0_1 \psi^0_2 b_m^*  + \psi^0_1\psi_2^{0*} d_m^* \nonumber \\
i\dot{b}_m &=& -\frac{1}{2} \Delta_H b_m + \frac{(m+S_2)^2}{2\rho^2} b_m  +V b_m -\mu_2 b_m \nonumber \\
&+& (|\psi^0_1|^2+2|\psi^0_2|^2) b_m +\psi_1^{0*} \psi^0_2 a_m + \psi^0_1\psi^0_2 c_m + \psi_2^{02} d_m \nonumber \\
i\dot{d}_m^* &=& -\frac{1}{2} \Delta_H d_m^* + \frac{(m-S_2)^2}{2\rho^2} d_m^*  +V d_m^* -\mu_2 d_m^* \nonumber \\
&+&(|\psi^0_1|^2+2|\psi^0_2|^2) d_m^* +\psi^0_1 \psi^0_2 a_m^* + \psi_2^{02} b_m^* + \psi_1^{0*}\psi_2^0 c_m^*.  \nonumber \\
\end{eqnarray}

Normal modes of the form $(a_m(t),c_m(t),b_m(t),d_m(t))^T=(a_m(0),c_m(0),b_m(0),d_m(0))^T\exp(\lambda t)$ lead to the following eigenvalue problems of the 4$\times$4 matrix $M$, with eigenvalues $\lambda$:
\begin{eqnarray}
M_{11}&=&i\left( \frac{1}{2} \Delta_H - \frac{(m+S_1)^2}{2\rho^2}  -V  +\mu_1-2|\psi^0_1|^2-|\psi^0_2|^2 \right), \nonumber \\
M_{12}&=&-i(\psi_1^{0})^2, \nonumber \\
M_{13}&=&-i\psi^0_1\psi_2^{0*}, \nonumber \\
M_{14}&=&-i\psi^0_1\psi^0_2, \nonumber \\
M_{21}&=&i(\psi_1^{0*})^2, \nonumber \\
M_{22}&=&i\left( -\frac{1}{2} \Delta_H + \frac{(m-S_1)^2}{2\rho^2}  +V  -\mu_1+2|\psi^0_1|^2+|\psi^0_2|^2 \right), \nonumber \\
M_{23}&=&i\psi_1^{0*}\psi_2^{0*}, \nonumber \\
M_{24}&=&i\psi_1^{0*}\psi^0_2, \nonumber \\
M_{31}&=&-i\psi_1^{0*}\psi^0_2, \nonumber \\
M_{32}&=&-i\psi^0_1\psi^0_2, \nonumber \\
M_{33}&=&-i\left( -\frac{1}{2} \Delta_H + \frac{(m+S_2)^2}{2\rho^2}  +V  -\mu_2+|\psi^0_1|^2+2|\psi^0_2|^2 \right), \nonumber \\
M_{34}&=&-i(\psi_2^{0})^2, \nonumber \\
M_{41}&=&i\psi_1^{0*}\psi_2^{0*}, \nonumber \\
M_{42}&=&i\psi^0_1\psi_2^{0*}, \nonumber \\
M_{43}&=&i(\psi_2^{0*})^2, \nonumber \\
M_{44}&=&i\left( -\frac{1}{2} \Delta_H + \frac{(m-S_2)^2}{2\rho^2}  +V  -\mu_2+|\psi^0_1|^2+2|\psi^0_2|^2 \right). \nonumber \\
\end{eqnarray}

One can prove using an orthogonal transformation and the symmetry of the $M$ matrix that the eigenvalues of the mode $-m$ are complex conjugates of the mode $m$, therefore, one only needs to compute eigenvalues of the non-negative $m$ modes. In the following calculations, we use $m=0,1,2,...,5$ as in Ref.~\cite{dss}. We apply the methods to the planar dark bright (PDB) and the spherical dark bright (SDB) solitary waves in this work. Note that the method works for a large class of solitons that are of interest including but not limited to the vortex-line-bright, vortex-ring-bright~\cite{stathis} and the skyrmion states~\cite{ruosteko}.


\end{document}